\RequirePackage{fix-cm}
\documentclass[smallextended]{svjour3}       
\smartqed  
\usepackage{graphicx}
\journalname{Hyperfine Interactions}
\begin{document}

\title{QCD symmetries and the $\eta$ and $\eta'$ in nuclei}

\author{Steven D. Bass}

\institute{S. D. Bass \at
        	Stefan Meyer Institute for Subatomic Physics,
		Austrian Academy of Sciences, \\
        	Boltzmanngasse 3,
		A 1090 Vienna,
		Austria 
}

\date{Received: date / Accepted: date}

\maketitle

\begin{abstract}
We discuss the role of QCD symmetries in understanding 
the $\eta$ and $\eta'$ mesons in nuclear media.
Recent results on the $\eta'$ mass in nuclei from the CBELSA/TAPS collaboration are very similar to the prediction of the Quark Meson Coupling model. 
\end{abstract}

\section{Introduction}

Recent progress in theoretical and experimental 
studies of the $\eta-$ and $\eta'-$ (as well as pion and kaon)
nucleus systems 
promises to yield valuable new information about
dynamical chiral and axial U(1) symmetry breaking 
in low energy QCD \cite{bt2013}.
With increasing nuclear density chiral symmetry is 
partially restored corresponding to a reduction in 
the values of the quark condensate and pion decay constant 
$f_{\pi}$ \cite{kienle,suzuki}.
This in turn leads to changes in the properties of hadrons 
in medium including the masses of the Goldstone bosons. 
While pions and kaons are would-be Goldstone bosons associated 
with chiral symmetry, the isosinglet $\eta$ and $\eta'$ mesons 
are too massive by about 300-400 MeV for them to be pure 
Goldstone states.
They receive extra mass from non-perturbative gluon dynamics associated with the QCD axial anomaly;
for recent reviews see \cite{bt2013,shore}.
How does this gluonic part change in nuclei ?
Medium modifications need to be understood 
self-consistently within the interplay of confinement,
spontaneous chiral symmetry breaking and axial U(1) dynamics.

The $\eta$- and $\eta'$-nucleon interactions are believed 
to be attractive corresponding to a reduced effective mass 
in the nuclear medium and the possibility that 
these mesons might form strong-interaction bound-states in nuclei.
For the $\eta$ one finds a sharp rise in the cross section at
threshold for $\eta$ production in both
photoproduction from $^3$He \cite{mamieta}
and in proton-deuteron collisions \cite{pdeta}
which may hint at a reduced $\eta$ effective mass 
in the nuclear medium.
Measurement of the $\eta'$-nucleus optical potential by 
the CBELSA/TAPS collaboration suggests that the effective 
$\eta'$ mass drops by about 40 MeV at nuclear matter density
\cite{nanova}.
For the pion and kaon systems one finds a small pion 
mass shift of order a few MeV in nuclear matter \cite{kienle}
whereas kaons are observed to experience an effective 
mass drop for the $K^-$ to about 270 MeV at two times 
nuclear matter density in heavy-ion collisions 
\cite{gillitzer,kaons}.
The same heavy-ion experiments also suggest the effective 
mass of anti-protons is reduced by about 100-150 MeV 
below their mass in free space \cite{gillitzer}.
The $\eta$-nucleon interaction is characterised by a strong coupling to the $S_{11}$(1535) nucleon resonance.
For example, $\eta$ meson production 
in proton nucleon collisions close to threshold is known 
to procede via a strong isovector exchange contribution 
with excitation of the $S_{11}(1535)$ \cite{pawela}.
Recent measurements of $\eta'$ production suggest a different
mechanism for this meson \cite{pawelb}. 
Experiments in heavy-ion collisions \cite{averbeck} and $\eta$ photoproduction from nuclei \cite{robig,yorita} suggest little modification of the $S_{11}(1535)$ excitation in-medium, 
though some evidence for the broadening of the $S_{11}$ 
in nuclei was reported in \cite{yorita}.

There is presently vigorous experimental 
\cite{nanova,elsa,pawelc,gsi,elsap,metagexa} 
and theoretical \cite{bt2013,etaA,hirenzaki,jido1,gal1} 
activity aimed at understanding 
the $\eta$ and $\eta'$ in medium and 
to search for evidence of possible
$\eta$ and $\eta'$ bound states in nuclei.
QCD inspired models of the $\eta$ and $\eta'$ nucleus 
systems are constructed with different selections of 
``good physics input'':
how they treat confinement, chiral symmetry and axial U(1) dynamics.
Here we focus on the 
Quark Meson Coupling model (QMC, for a review see \cite{qmc}). 
In the QMC model medium modifications are calculated at
the quark level through coupling of the light quarks in 
the hadron to the scalar isoscalar $\sigma$ 
(and also $\omega$ and $\rho$) mean fields in the nucleus.
Possible binding energies and the in-medium masses of 
the $\eta$ and $\eta'$ are sensitive to the flavour-singlet component in the mesons and hence to the non-perturbative glue associated with axial U(1) dynamics \cite{etaA}.

Meson masses in nuclei are determined from the meson nucleus optical potential and the scalar induced contribution 
to the meson propagator evaluated at zero three-momentum, 
${\vec k} =0$, in the nuclear medium.
Let $k=(E,{\vec k})$ and $m$ denote the four-momentum and mass of the meson in free space.
Then, one solves the equation
\begin{equation}
k^2 - m^2 = {\tt Re} \ \Pi (E, {\vec k}, \rho)
\end{equation}
for ${\vec k}=0$
where $\Pi$ is the in-medium $s$-wave meson self-energy.
Contributions to the in medium mass come from coupling to the scalar 
$\sigma$ field in the nucleus in mean-field approximation,
nucleon-hole and resonance-hole excitations in the medium.
For ${\vec k}=0$, $k^2 - m^2 \sim 2 m (m^* - m)$ 
where $m^*$ is the effective mass in the medium.
The mass shift $m^*-m$
is the depth or real part of the meson nucleus optical potential.
The imaginary part of the potential measures the width of 
the meson in the nuclear medium.
The $s$-wave self-energy can be written as \cite{ericson}
\begin{equation}
\Pi (E, {\vec k}, \rho) \bigg|_{\{{\vec k}=0\}}
=
- 4 \pi \rho \biggl( { b \over 1 + b \langle {1 \over r} \rangle } \biggr) .
\end{equation}
Here $\rho$ is the nuclear density,
$
b = a ( 1 + {m \over M} )
$
where 
$a$ is the meson-nucleon scattering length, 
$M$ is the nucleon mass and
$\langle {1 \over r} \rangle$ is the inverse correlation length,
$\langle {1 \over r} \rangle \simeq m_{\pi}$ 
for nuclear matter density.
Attraction corresponds to positive values of $a$.
The denominator in Eq.(2) is the 
Ericson-Ericson-Lorentz-Lorenz double scattering correction.

Meson mass shifts in medium can be investigated 
through studies of excitation functions in photoproduction 
experiments from nuclear targets
and through searches for possible meson bound states in nuclei. 
In photoproduction experiments
the production cross section is enhanced with the lower effective meson mass in the nuclear medium. 
When the meson leaves the nucleus it returns on-shell 
to its free mass with the energy budget conserved at
the expense of the kinetic energy so that excitation functions
and momentum distributions can provide essential clues to the
meson properties in medium \cite{metag}.
Using this physics a first (indirect) estimate of the $\eta'$ 
mass shift has recently been deduced 
by the CBELSA/TAPS Collaboration \cite{nanova}.
The $\eta'$-nucleus optical potential 
$V_{\rm opt} = V_{\rm real} + iW$
deduced from these photoproduction experiments is 
\begin{eqnarray}
V_{\rm real} (\rho_0)
= m^* - m 
&=& -37 \pm 10 (stat.) \pm 10 (syst.) \ {\rm MeV}
\nonumber \\ 
W(\rho_0) &=& -10 \pm 2.5 \ {\rm MeV}
\end{eqnarray}
at nuclear matter density $\rho_0$.
In this experiment the average momentum of the produced 
$\eta'$ was 1.1 GeV and the mass shift was measured in
production from a carbon target.
The mass shift, Eq.(3), is very similar 
to the expectations of the Quark Meson Coupling model, 
see below.
If substituted into Eq.(2) with the Ericson-Ericson 
denominator switched off, 
then one finds an effective scattering length with
real part of 0.5 fm.
The COSY-11 collaboration have recently determined 
the $\eta'$-nucleon scattering length in free space
to be
\begin{eqnarray}
\nonumber
\mathrm{Re}(a_{\eta' p}) &=&  0~\pm~0.43~\mathrm{fm}
\\
\mathrm{Im}(a_{\eta' p}) &=& 0.37^{~+0.40}_{~-0.16}~\mathrm{fm}\end{eqnarray}
from studies of the final state interaction in $\eta'$ production in proton-proton collisions close to threshold 
\cite{eryk2014}.
Theoretical models in general prefer a 
positive sign for the real part of $a_{\eta' p}$.

New experiments are planned to look for possible $\eta'$ 
bound states in carbon using the (p, d) reaction at GSI 
\cite{gsi} and in photoproduction at ELSA \cite{elsap}.
The small $\eta'$ width in nuclei $20 \pm 5.0$ MeV at 
nuclear matter density in Eq.(3) was extracted 
from measurements of the transparency ratio for $\eta'$ photoproduction from nuclear targets \cite{elsa}
and suggests the possibility of relatively narrow bound 
$\eta'$-nucleus states accessible to experiments.
For clean observation of a bound state one needs the real part 
of the optical potential to be much bigger than the imaginary part.
COSY searches are focussed on 
possible $\eta$ bound states in $^3$He and $^4$He 
\cite{pawelc}.

\section{QCD symmetries and the $\eta$ and $\eta'$}

Spontaneous chiral symmetry breaking in QCD 
induces an octet of Goldstone bosons associated 
with SU(3) and also 
(before extra gluonic effects in the singlet channel)
a flavour-singlet Goldstone boson.
The mass squared of these Goldstone bosons 
is proportional to the current mass of their valence quarks.
While the pion and kaon fit well in this picture,
to understand the 
isosinglet $\eta$ and $\eta'$ masses
one needs extra mass in the flavour-singlet channel 
associated 
with non-perturbative topological gluon configurations 
\cite{shore,crewther}, 
related perhaps to confinement \cite{ks} or instantons 
\cite{thooft}.
The gluonic mass term ${\tilde m}^2_{\eta_0}$ satisfies the Witten-Veneziano mass formula 
\cite{witten,venez}
\begin{equation}
m_{\eta}^2 + m_{\eta'}^2 = 2 m_K^2 + {\tilde m}_{\eta_0}^2 
\end{equation}
and has a rigorous interpretation in terms of 
the QCD Yang-Mills topological susceptibility. 
SU(3) breaking generates mixing between 
the octet and singlet states which, together with the gluonic mass contribution, yields the massive $\eta$ and $\eta'$ bosons.
Phenomenological studies of various decay processes give a value 
for the $\eta$-$\eta'$ mixing angle between $-15^\circ$ and 
$-20^\circ$ \cite{gilman}.
In the OZI limit of no gluonic mass term
the $\eta$ would be approximately an isosinglet light-quark state
(${1 \over \sqrt{2}} | {\bar u} u + {\bar d} d \rangle$)
with mass $m_{\eta} \sim m_{\pi}$
degenerate with the pion and
the $\eta'$ would be a strange-quark state $| {\bar s} s \rangle$
with mass $m_{\eta'} \sim \sqrt{2 m_{\rm K}^2 - m_{\pi}^2}$,
mirroring the isoscalar vector $\omega$ and $\phi$ mesons.

The gluonic mass term is related to the QCD axial anomaly 
in the divergence of the flavour-singlet axial-vector current.
While the non-singlet axial-vector currents are partially conserved (they have just mass terms in the divergence), the singlet current
$
J_{\mu 5} = \bar{u}\gamma_\mu\gamma_5u
+ \bar{d}\gamma_\mu\gamma_5d + \bar{s}\gamma_\mu\gamma_5s 
$
satisfies the anomalous divergence equation 
\begin{equation}
\partial^\mu J_{\mu5} = 6 Q
+ \sum_{k=1}^{3} 2 i m_k \bar{q}_k \gamma_5 q_k 
\end{equation}
where 
$
Q = \partial^{\mu} K_{\mu}
= {\alpha_s \over 8 \pi} G_{\mu \nu} {\tilde G}^{\mu \nu}
$
is the topological charge density.
The integral over space $\int \ d^4 z \ Q = n$ measures the 
gluonic winding number \cite{crewther} 
which is an integer for (anti-)instantons and which vanishes 
in perturbative QCD.

$\eta$-$\eta'$ mixing means that non-perturbative glue through axial U(1) dynamics plays an important role in both the $\eta$ and $\eta'$ mesons and their interactions.
The anomalous glue that generates the large $\eta$ and $\eta'$ 
masses also drives OZI violating $\eta$ and $\eta'$ production 
and decay processes [33-37] and 
enters in the $\eta'$-nucleon interaction \cite{bass99}
that has been the subject of 
vigorous experimental investigation at COSY \cite{pawelcosy11}.
The QCD axial anomaly also 
plays an important role in interpetation of the nucleon's
flavour-singlet axial-charge (or ``quark spin content'')
measured in polarised deep inelastic scattering and
associated with the proton spin puzzle \cite{SBrmp1,SBrmp2}.

Within the low energy effective chiral Lagrangian for QCD 
the gluonic mass term ${\tilde m}_{\eta_0}^{2}$ 
is introduced via a flavour-singlet
potential involving the topological charge density $Q$
which is constructed so that the Lagrangian also reproduces 
the axial anomaly \cite{vecca}.
In this approach the medium dependence of 
${\tilde m}_{\eta_0}^2$ 
is introduced through coupling to the $\sigma$ 
(correlated two-pion) mean-field in the nucleus
through the interaction term
$
{\cal L}_{\sigma Q} = g_{\sigma Q} \ Q^2 \ \sigma
$
where 
$g_{\sigma Q}$ denotes coupling to the $\sigma$ mean field.
One finds the gluonic mass term decreases in-medium
${\tilde m}_{\eta_0}^{*2} < {\tilde m}_{\eta_0}^2$ 
independent of the sign of $g_{\sigma Q}$.
The medium acts to partially neutralise 
axial U(1) symmetry breaking by gluonic effects \cite{etaA}.

As a second interesting application of the QCD effective Lagrangian approach,
the OZI violating interaction
$\lambda Q^2 \partial_{\mu} \pi_a \partial^{\mu} \pi_a$
with $\pi_a$ the pseudoscalar Goldstone fields
is needed to generate the leading (tree-level)
contribution to the decay $\eta' \rightarrow \eta \pi \pi$
\cite{veccb}.
When iterated in the Bethe-Salpeter equation for $\eta' \pi$
rescattering
this interaction yields a dynamically generated 
resonance with quantum numbers $J^{PC} = 1^{-+}$ 
and mass about 1400 MeV.
The generation of this state is mediated 
by the OZI violating coupling of the $\eta'$ \cite{bassmarco}.
One finds a possible dynamical interpretation of the 
light-mass $1^{-+}$ exotics observed in experiments at BNL \cite{exoticb} and CERN \cite{exoticc}.
This OZI violating interaction will also contribute to 
higher $L$ odd partial waves with quantum numbers $L^{-+}$.
These states are particularly interesting because 
the quantum numbers $1^{-+}, 3^{-+}, 5^{-+}$...
are inconsistent with a simple quark-antiquark bound state.
The COMPASS experiment at CERN has recently measured 
exclusive production of $\eta' \pi^-$ and $\eta \pi^-$ 
in 191 GeV $\pi^-$ collisions on a hydrogen target
\cite{compassexotic}.
They find the interesting result that
$\eta' \pi^-$ production 
is enhanced relative to $\eta \pi^-$ production 
by a factor of 5-10 in the exotic $L=1,3,5$ partial waves 
with quantum numbers $L^{-+}$ in the inspected invariant 
mass range up to 3 GeV. No enhancement was observed in the
even $L$ partial waves.

\section{The $\eta$ and $\eta'$ in nuclei}

The physics of the $\eta$ and $\eta'$ in medium has been
investigated by Bass and Thomas \cite{etaA} within 
the Quark Meson Coupling model \cite{qmc,guichon,etaqmc}
taking into account $\eta$-$\eta'$ mixing and 
the flavour-singlet component in these mesons.
In these calculations the large $\eta$ and $\eta'$ 
masses are used to motivate 
taking an MIT Bag description for the meson wavefunctions.
Gluonic topological effects are understood to be ``frozen in'',
meaning that they are only present implicity through the masses
and mixing angle in the model.
The in-medium mass modification comes from coupling the light 
(up and down) quarks and antiquarks in the meson wavefunction 
to the scalar $\sigma$ mean-field in the nucleus 
working in mean-field approximation \cite{qmc}.
The coupling constants in the model for the coupling of 
light-quarks to the $\sigma$ (and $\omega$ and $\rho$) 
mean-fields in the nucleus are adjusted to fit the 
saturation energy and density of
symmetric nuclear matter and the bulk symmetry energy.
The strange-quark component of the wavefunction does not couple to the $\sigma$ field and $\eta$-$\eta'$ mixing is readily built into the model.
Gluon fluctuation and centre-of-mass effects are assumed to be independent of density.
The model results for the meson masses in medium and the real 
part of the meson-nucleon scattering lengths are shown 
in Table 1 for different values of the $\eta$-$\eta'$ 
mixing angle which is taken to be density independent.

With an $\eta$-$\eta'$ mixing angle of $-20^\circ$ 
the QMC prediction for the $\eta'$ mass in medium at nuclear matter density is 921 MeV, that is a mass shift of $-37$ MeV. 
This value is in excellent agreement with the mass shift 
$-37 \pm 10 \pm 10$ MeV deduced from photoproduction data 
\cite{nanova}.
Mixing increases the octet relative to singlet component in
the $\eta'$, reducing the binding through increased strange
quark component in the $\eta'$ wavefunction.
Without the gluonic mass contribution the $\eta'$ 
would be a strange quark state after $\eta$-$\eta'$ mixing.
Within the QMC model there would be no coupling to the 
$\sigma$ mean field and no mass shift so that any observed mass shift is induced by glue associated with the QCD axial anomaly that generates part of the $\eta'$ mass.

Increasing the flavour-singlet component in the $\eta$ at 
the expense of the octet component gives more attraction, 
more binding and a larger value of the $\eta$-nucleon scattering length, $a_{\eta N}$. 
$\eta$-$\eta'$ mixing with the phenomenological mixing angle 
$-20^\circ$ leads to a factor of two increase 
in the mass-shift and 
in the scattering length obtained in the model
relative to the prediction for a pure octet $\eta_8$.
This result may explain why values of $a_{\eta N}$ 
extracted from phenomenological fits to experimental 
data where the $\eta$-$\eta'$ mixing angle is unconstrained 
\cite{wycech} 
give larger values 
(with real part about 0.9 fm)
than those predicted 
in theoretical coupled channels models where the $\eta$ 
is treated as a pure octet state \cite{etaweise,etaoset}.

\begin{table}[t!]
\begin{center}
\caption{
Physical masses fitted in free space, the bag masses in medium at normal nuclear-matter density, $\rho_0 = 0.15$ fm$^{-3}$, 
and corresponding effective meson-nucleon scattering lengths.
The values of ${\tt Re} a_{\eta}$ are obtained with 
the Ericson-Ericson denominator turned-off (since we work in mean-field approximation).}

\label{bagparam}
\begin{tabular}[t]{c|lll}
\hline
&$m$ (MeV) 
& $m^*$ (MeV) & ${\tt Re} a$ (fm)
\\
\hline
$\eta_8$  &547.75  
& 500.0 &  0.43 \\
$\eta$ (-10$^o$)& 547.75  
& 474.7 & 0.64 \\
$\eta$ (-20$^o$)& 547.75  
& 449.3 & 0.85 \\
$\eta_0$  &      958 
& 878.6  & 0.99 \\
$\eta'$ (-10$^o$)&958 
& 899.2 & 0.74 \\
$\eta'$ (-20$^o$)&958 
& 921.3 & 0.47 \\
\hline
\end{tabular}
\end{center}
\end{table}

For baryons in symmetric nuclear matter the QMC model
predicts an effective proton mass about 755 MeV at nuclear matter density \cite{qmc}.
The $S_{11}$ is interpreted in quark models as a 
3-quark state $(1s)^2(1p)$. 
In QMC one finds an excitation energy of $\sim 1544$ MeV, consistent with observations,
with the scalar attraction compensated by repulsion from
coupling to the $\omega$ mean-field to give the excitation energy \cite{etaA}.
Small mass shift is also found in coupled channels models 
where the $S_{11}$ is instead interpreted as a $K \Sigma$ 
quasi-bound state, with the $\eta$ instead treated as a pure octet state \cite{kaiser}.

For the $\eta'$ in medium, larger mass shifts, downwards 
by up to 80-150 MeV, were found in recent 
Nambu-Jona-Lasinio model calculations (without confinement) \cite{hirenzaki} and in linear sigma model calculations 
(in a hadronic basis) \cite{jido1} which also suggest a 
rising $\eta$ effective mass at finite density. 
A chiral coupled channels calculation performed with
possible $\eta'$-nucleon scattering lengths with real 
part between 0 and 1.5 fm is reported in \cite{osetetaprime}.

\section{Outlook}

Medium modifications of hadron properties are determined 
by chiral and flavour symmetries in QCD.
The $\eta$ and $\eta'$ are sensitive to flavour-singlet 
axial U(1) degrees of freedom.
QCD inspired models including confinement, chiral and axial 
U(1) dynamics yield a range of predictions 
for the $\eta$ and $\eta'$ mass shifts in nuclei 
and the corresponding meson-nucleon scattering lengths.
The QMC prediction for the $\eta'$ mass shift is 
very similar to the recent value determined by CBELSA/TAPS 
from photoproduction experiments.
The model value for the real part of the $\eta$-nucleon scattering length is also close to values extracted from phenomenological fits to low-energy scattering data.
New data on the $\eta$ and $\eta'$ in nuclei and possible
bound states are expected soon from running and planned experiments at COSY, ELSA and GSI, and will help 
further constrain our understanding of axial U(1) dynamics
in low-energy QCD.


\begin{acknowledgements}

I thank M. Faessler, V. Metag, P. Moskal and K. Suzuki 
for helpful discussions.
I thank the organisers for the invitation to talk at 
this stimulating meeting.
The research of SDB is supported 
by the Austrian Science Fund, FWF, through grant P23753.
Part of this work was supported by the LEANNIS networking activity within the European Community Research 
Infrastructure Integrating Activity 
``Study of Strongly Interacting Matter'' 
(HadronPhysics3 (HP3) Contract No. 283286) 
under the Seventh Framework Programme of EU.

\end{acknowledgements}



\end{document}